\documentclass{KapProc} 

\let\footnote\savefootnote
\let\footnotetext\savefootnotetext 
\let\footnoterule\savefootnoterule 

\kluwerbib

\startauthorindex

\begin{document}


\articletitle[]{Further Detections of OH Masers \\
in Carbon Stars with Silicate \\Features}

\author{M. Szymczak}
\affil{Toru\'n Centre for Astronomy, Nicolaus Copernicus University,
Toru\'{n}, Poland} 
\author{R. Szczerba}
\affil{Nicolaus Copernicus Astronomical Center, PAS, Toru\'{n}, Poland}
\author{P.S. Chen}
\affil{Yunnan Astronomical Observatory, CAS, Kunming, P.R. China }

\chaptitlerunninghead{OH Masers in Silicate Carbon Stars}


\begin{abstract}
A sample of J--type carbon stars was searched for OH maser emission. The new 
detection of three OH lines towards two silicate carbon 
stars is reported. In V778\,Cyg, previously known as the main--lines (1665 
and 1667\,MHz) maser source, the satellite 1612\,MHz emission was discovered 
while in NSV\,2814 the main OH lines were detected. The presence of OH maser 
lines confirms the former suggestion that oxygen--rich material is located in 
the vicinity ($\approx$\,$10^{15-16}$\,cm) of silicate carbon stars. 
\end{abstract}

\section{Introduction}
\label{intr}

The silicate emission features at about 10 and 18\,$\mu$m are characteristic 
for oxygen--rich dust envelopes. However, these features were also discovered
in some optically classified carbon stars 
(Little-Marenin\,\cite{littlemarenin86}; 
Willems \& de Jong\,\cite{willems86}), named later silicate carbon stars. 
Furthermore, many silicate carbon stars are 
recognized as $^{13}$\,C-rich (J-type) carbon stars 
(Lambert et al.\,\cite{lambert90}; Lloyd Evans\,\cite{lloydevans90} and 
references therein). 
The most recent compilation of 22 known and suspected silicate carbon stars 
is presented by Chen et al.\,(\cite{chen99}). 
Note also the discovery of the first extragalactic silicate carbon star 
(IRAS\,04496$-$6958) in the Large Magellanic Cloud 
(Trams et al.\,\cite{trams99}). The detection of silicate emission from 
some (silicate) carbon stars suggests that their relatively close surroundings 
contain oxygen--based dust, in spite of
their photospheric chemical composition which shows C/O\,$>$\,1. An additional
argument for the existence of oxygen--rich material in the vicinity
of these stars comes from the detection of water and hydroxyl maser 
lines towards some of them
(see Little-Marenin et al.\,\cite{littlemarenin94} and Engels\,\cite{engels94}
and references therein). To date H$_2$O masers were found in four (J--type) 
silicate carbon stars and only one of them (V778\,Cyg) exibits OH maser 
emission in the main--lines. In addition 1612\,MHz maser emission was detected 
in FJF\,270 (te Lintel Hekkert\,\cite{telintelhekkert91}). 
Here we report the results of a high sensitivity search for OH 
emission towards 14 J--type carbon stars including 7 known silicate carbon 
stars.

\section{Observations}

The observations were performed with the Nan\c{c}ay radio telescope on April 
6--13, 1999. The HPBW was $3.5^{\prime}$ in right ascension by $18^{\prime}$ in
declination. A dual channel receiver was used. The system temperature was 
about 50\,K. The ratio of flux density to antenna temperature was 
1.1\,Jy\,K$^{-1}$ at $0^{\circ}$ declination. The 1024--channel 
autocorrelation spectrometer was split into 4 banks, each covering a bandwidth
of 0.2\,MHz, for the three OH lines. The 1612\,MHz satellite line was observed
in both circular polarizations while the 1665 and 1667\,MHz main--lines 
were observed in left and right circular polarization, respectively. The 
spectra were taken in frequency switching mode with channel spacings of 0.28 
and 0.29\,km\,s$^{-1}$ at the main--lines and the satellite line, 
respectively. Upper flux density limit for the non--detections was 
80\,mJy (3$\sigma$). W\,12 was observed to provide the flux density 
calibration which was accurate to about 10\%.

\section{Results and Discussion}

Our program stars are listed in Table\,1 with the designation from the General 
Catalogue of Cool Galactic Carbon Stars 
(CCGCS, Stephenson\,\cite{stephenson89})
given in the first column. Known silicate carbon stars are marked in boldface.
Previously detected H$_2$O and/or OH maser lines are marked by a `+' sign 
while our new detections are depicted by a `++' sign. OH masers were 
found in NSV\,2814 (main--lines) and in V778\,Cyg (satellite line). Note that
the 1612\,MHz maser line towards V778\,Cyg has not been detected either by 
Barnbaum et al.\,(\cite{barnbaum91}) nor by 
Little-Marenin et al.\,(\cite{littlemarenin94}).

\begin{table}[ht]
\caption{List of observed stars.}
 \centering
 \begin{tabular}{l l c c c c}\hline
 CCGCS & other name & OH 1612 & OH 1665 & OH 1667 & H$_2$O \\
\hline
{\bf 1158} & {\bf NSV 2814}  & $-$ &  ++ &  ++  &  + \\  
{\bf 1653} & {\bf BM Gem} & $-$ & $-$ & $-$ & $-$ \\
1682 & & $-$ & $-$ & $-$ & + \\
1698 & & $-$ & $-$ & $-$ & + \\
1921 & & $-$ & $-$ & $-$ & $-$\\
2028 & & $-$ & $-$ & $-$ & $-$\\
2156 & MT Hya & $-$ & $-$ & $-$ & $-$\\
3066 & HD 100764 & $-$ & $-$ & $-$ & $-$\\
{\bf 3935} & {\bf FJF 270} & + & $-$ & $-$ & $-$ \\
{\bf 4222} & {\bf NC 83} & $-$ & $-$ & $-$ & $-$\\
{\bf 4923} & {\bf V778 Cyg} & ++ &  + &  +  &   +  \\
{\bf 5526} & {\bf MQ Cyg} & $-$ & $-$ & $-$ & $-$\\
5548 & V1415 Cyg & $-$ & $-$ & $-$ & + \\
{\bf 5848} & {\bf EU And} & $-$ & $-$ & $-$ & + \\
\hline
 \end{tabular}
 \end{table}
\begin{figure}[ht]
\vskip3.15in
\hskip-0.2in
\includegraphics{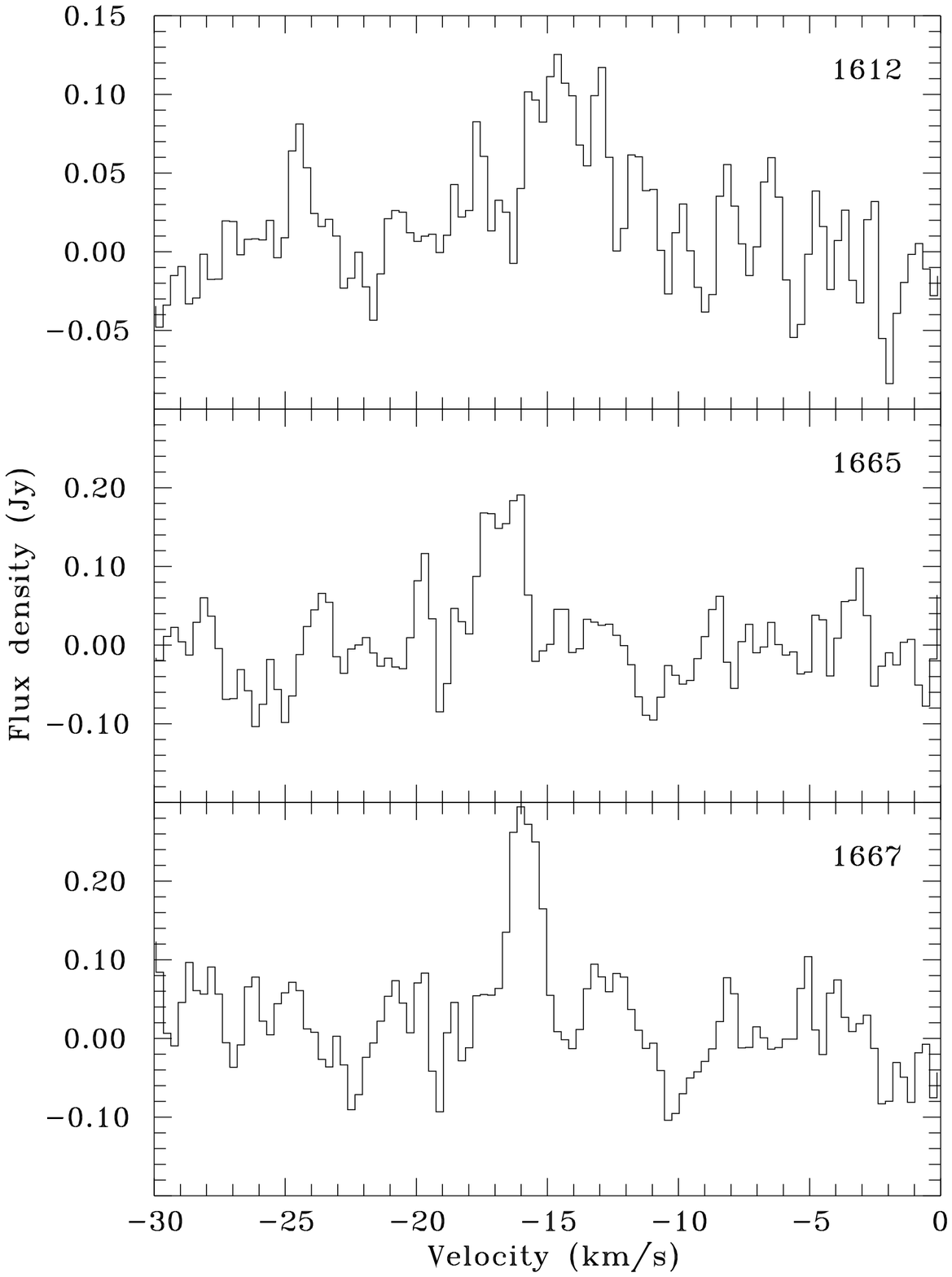}
\includegraphics{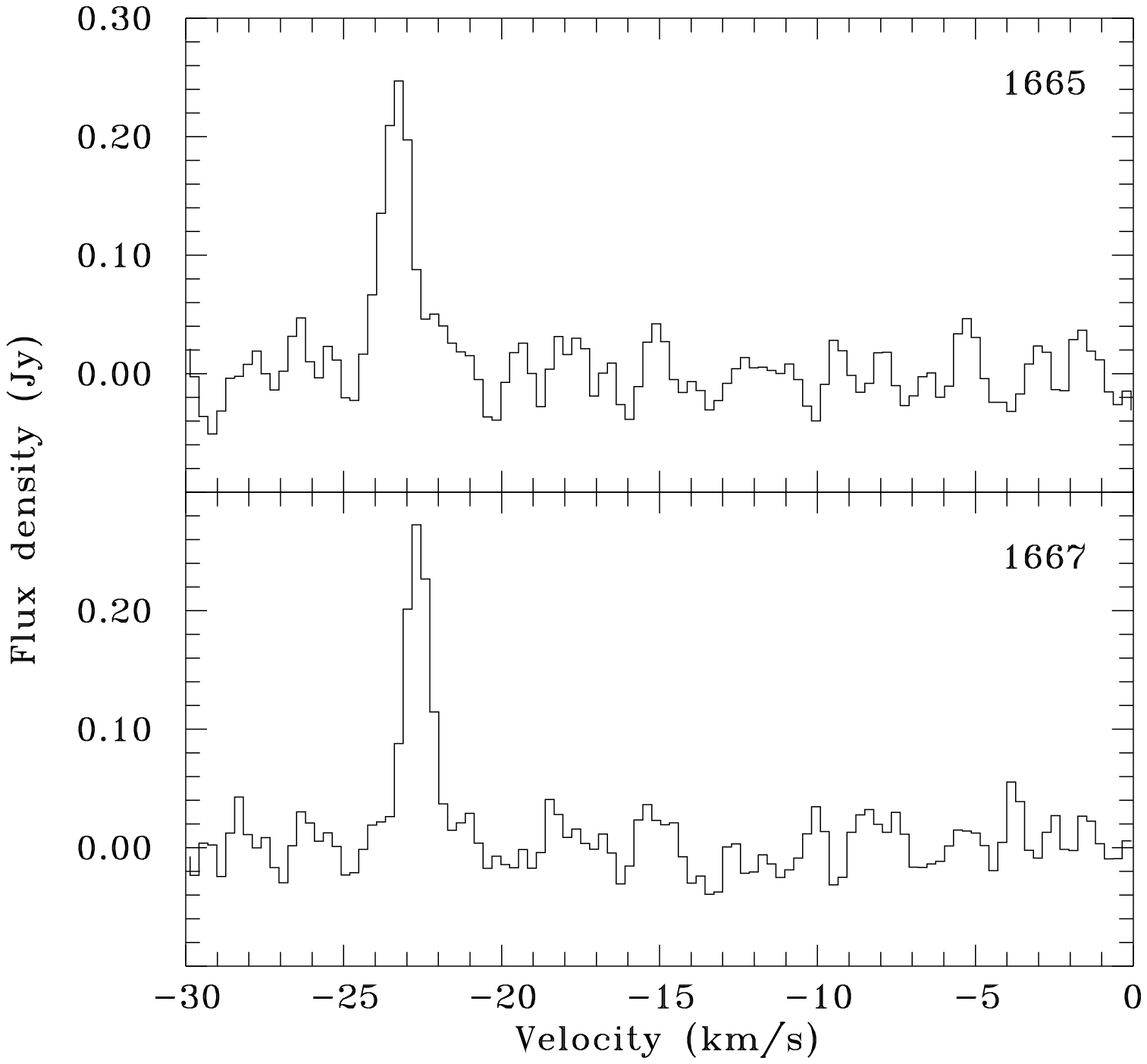}
\vskip-1.2in
\narrowcaption{OH maser spectra of V778\,Cyg (left) and NSV\,2814 (right). 
Transition frequencies in MHz are shown.}
\end{figure}

The spectra are shown in Fig.\,1 where, for the sake of comparison, 
re--observed OH main--lines from V778\,Cyg are presented. 
In V778\,Cyg the 1667\,MHz feature at $-$16.0\,km\,s$^{-1}$ has a peak
flux density of 0.30\,Jy, which is a factor of 1.5 higher than the 1665\,MHz
peak of 0.19\,Jy at $-$16.9\,km\,s$^{-1}$. Within the measurement errors
the intensities and radial velocities of these lines are the same as 
observed 12 years ago (Little--Marenin et al.\,\cite{littlemarenin88}).
The 1612\,MHz feature of 0.12\,Jy peaks at $-$14.4\,km\,s$^{-1}$. This
indicates that the 1612\,MHz emission is slightly redshifted relatively 
to the OH main--lines but coincides with a $-$14\,km\,s$^{-1}$ component of
the H$_2$O maser (Engels \& Leinert\,\cite{engelsleinert94}).

In NSV\,2814 the peak intensities of the 1665 and 1667\,MHz are 0.24 
and 0.27\,Jy, respectively. The velocities of OH features at $-$23.5 and
$-$22.8\,km\,s$^{-1}$ are very different from the H$_2$O maser velocity 
of $-$13\,km\,s$^{-1}$ detected about 7 years ago (Engels\,\cite{engels94}).
Although temporal variations may affect the maser profiles, the difference
in velocities of the OH and H$_2$O maser features of about 10\,km\,s$^{-1}$ is
quite high for an AGB object. This may suggest that each maser is associated
with a different object. Little-Marenin et al.\,(\cite{littlemarenin94}) 
noted that CCGCS\,1158 may be incorrectly associated with the IRAS source. 
Furthermore, the intensity ratio of 1667/1665\,MHz transitions of about 1.1 
suggests optically thin LTE conditions, so that the masers may arise in the 
intervening interstellar medium. The case of NSV\,2814 needs further careful
studies.  

In spite of almost 15 years of debate, the formation of silicate carbon
stars still remains controversial. A stability of the silicate 
feature (Yamamura et al.\,\cite{yamamura00}) and 
OH main--lines (this work) in V778 Cyg makes the scenario of {\it fast} 
transition from M--type AGB stars into carbon stars difficult to reconcile 
with the lifetime of silicate emission from an expanding detached shell
(see e.g. Yamamura et al.\,\cite{yamamura00}).
A plausible scenario may be some kind of long--lived reservoirs of 
oxygen--rich material in a binary system 
(e.g. Lloyd Evans\,\cite{lloydevans90}
and Jura \& Kahane\,\cite{jura99}) with (possibly) a main sequence 
secondary (e.g. Lambert et al.\,\cite{lambert90}
and Yamamura et al.\,\cite{yamamura00}).
Silicate carbon stars seem to be related to the {\it normal} J--type carbon 
stars in the sense that both groups show H$_2$O maser emission 
(Engels\,\cite{engels94}) and that the mechanism responsible for 
their low $^{12}$C/$^{13}$C ratios might be the same 
(Ohnaka \& Tsuji\,\cite{ohnaka99}).

\begin{acknowledgments}
The authors acknowledge support from grant No.\,2.P03D.002.13 of the Polish 
State Committee for Scientific Research.
\end{acknowledgments}

\end{document}